# Spectral homogeneity cross frequencies can be a quality metric for the large-scale resting EEG preprocessing

Shiang Hu[1*], Jie Ruan[1], Nicolas Langer[2], Jorge Bosch-Bayard[3,4], Zhao Lv[1], Dezhong Yao[3], Pedro Antonio Valdes-Sosa[3,4]

[1]Anhui Province Key Laboratory of Multimodal Cognitive Computation, School of Computer Science and Technology, Anhui University, 111 Jiulong Road, Hefei, 230601, China

[2]Methods of Plasticity Research, Department of Psychology, University of Zurich, Zurich, Switzerland

[3]The Clinical Hospital of Chengdu Brain Science Institute, MOE Key Lab for Neuroinformation, School of Life Science and Technology, University of Electronic Science and Technology of China, Chengdu, China

[4]Cuban Center for Neurocience, La Habana, Cuba

Corresponding author: shu@ahu.edu.cn

# Abstract

The brain projects require the collection of massive electrophysiological data, aiming to the longitudinal, sectional, or populational neuroscience studies. Quality metrics automatically label the data after centralized preprocessing. However, although the waveforms-based metrics are partially useful, they may be unreliable by neglecting the spectral profiles. Here, we detected the phenomenon of parallel log spectra (PaLOS) that the scalp EEG power in the log scale were parallel to each other from 10% of 2549 HBN EEG. This phenomenon was reproduced in 8% of 412 PMDT EEG from 4 databases. We designed the PaLOS index (PaLOSi) to indicate this phenomenon by decomposing the cross-spectra at different frequencies into the common principal component spaces. We found that the PaLOS biophysically implied a prominently dominant dipole in the source space which was implausible for the resting EEG. And it may be practically resulted from excessive preprocessing. Compared with the 1966 normative EEG cross-spectra, the HBN and the PMDT EEG with PaLOS presented generally much higher electrode pairwise coherences and higher similarity of coherence-based network patterns, which went against the known frequency dependent characteristic of coherence networks. We suggest the PaLOSi should lay in the range of 0.4-0.7 for large resting EEG quality assurance.

## 1. Introduction

### 1.1 The general electrophysiology pipelines

The emergence of large scale Magneto and Electroencephalographic (MEEG) recordings provides the chance to study the populational norm related to aging and the mechanism of brain functions [1–3]. To understand the large scale MEEG recordings, many electrophysiology pipelines are developed for automatic batch processing. The core modules are automatic artifact removal, automatic forward modelling, automatic source estimation, automatic spectral connectivity analysis and automatic statistical analysis. The brain signals are always mixed with various interferences and biological noise, making the artifact removal to be the indispensable prerequisite. Automatic artifact removal is challenging because there is no common standard across subjects what the clean EEG should look like.

### 1.2 Description of artifact removal pipeline

Many artifact-removal pipelines have been developed, with the name of preprocessing pipelines, meant to clean the artifacts in the preprocessing stage. By the assumption of statistical independence, independent component analysis (ICA) is used to separate the raw recordings into different components, each of which corresponds to a topographical distribution. The early work to integrate ICA as tool or pipeline is EEGLAB [4,5]; later on, the typical are FASTER [6], ADJUST [7], MARA [8], PREP [9]; and the recent are Autoreject [10], APP [11], HAPPE [12], BEAPP [13], Automagic [14], ICLABEL [15], MADE [16], EEG-IP-L [17]. The mainstay of the above MEEG artifact-removal pipelines is ICA. Particularly, MARA and ICLABEL are the IC classifiers using the machine learning approach; meanwhile, EEG-IP-L aims to objectively retain data after annotation and ICA by interactively quality control.

Most preprocessing pipelines include bad channel detection and interpolation, filtering, linear noise removal, and automatic identification of bad independent components. However, only Automagic, CTAP, and HAPPE provide quality metrics. Automagic offers a user-friendly interface where users can simply specify the input data folder, the output folder for preprocessed results, and configure preprocessing parameters for batch processing. Additionally, Automagic allows users to save the data after each preprocessing step for inspection.

## 1.3 Propose the question of quality control

MEEG preprocessing pipelines are the essential first step before any analysis occurs. Preprocessing comprises of multiple steps, such as re-referencing, filtering, line noise removal, detrending, bad channels/epochs/trials detection, artifact components removal, segmentation, and Bad channel interpolation, etc. Our emphasis here will be on artifact detection and correction. For example, many efforts have been put to the classification and selection of Independent Components with the aim to automatically identify and discard the artifact components. To illustrate, CORRMAP calculated the correlation between the inverse weights of ICs and the templates to automatically identify the ICs [18]; MARA employed multiple features classifier to identify the artifact components [8]; SASICA provided interactive plots guiding for improving the manual selection of ICs by expert supervision [19]. ICA is thus the mainstay but not the whole body of artifact removal pipelines.

Although pipelines are rapidly evolving, there is still no metric to control the degree of artifact removal. The central point is how to balance the maximal rejection of all kinds of unwanted 'noise' from the raw recordings against the maximal retention of brain signals during the preprocessing in order not to affect mining the brain mechanism, that is, the problem of quality control in the large-scale EEG preprocessing where the expert supervision becomes impossible. We note that the EEG content that is removed in the preprocessing is not only the artifact components found by ICA but also the cumulative effect of each step in the pipeline.

## 1.4 Description of existing quality metric and point out the problems

The quality metrics proposed in the existing studies are the statistical thresholding from the temporal aspect features of the cleaned EEG waveforms. Those are the peak to peak threshold [10], the ratio of bad channels/epochs/segments, the number of rejected components, the ratio of timepoints/channels of high amplitude/variance [14,20], the ratio of interpolated channels/components and the percentage of retained variance after ICA back-projection [12], and the median/standard deviation for outlier estimation in the interactive quality control review after data annotation [17].

In terms of the percentage of retained variance, this metric shows how much variance is retained after projecting back the non-artifactual (clean) ICs. Note that retained variance is influenced by how clean the data are when entering the ICA of MARA/ICLabel. Therefore, more retained variance does not per se indicate less noisy data. If MARA is used without a preceding EOG regression, substantially more variance is removed from the data in the bad datasets (on average 52% retained variance in good datasets, 27% retained variance in bad datasets) whereas this difference is substantially smaller when MARA is used after EOG regression (on average 52% retained variance in good datasets, 42% retained variance in bad datasets). This indicates that the EOG regression and MARA may to some extent remove similar artifacts.

Seemingly, the current available quality metrics are to quantify statistical deviation of the preprocessed recordings from the good-looking waveform. The motiving concern for this paper is whether pursing the good-looking waveforms will result in excessive artifact removal and the loss of brain signals, and how to control this.

## 1.5 The EEG spectra

Controlling the quality of preprocessing is of great importance to the study of neural oscillation and frequency domain source connectivity analysis. The existing quality metrics are mostly the descriptive statistics of the temporal waveforms. Of neurophysiology relevance is the transformed representation in the frequency domain, such as the power spectra, cross spectra, coherence, and connectivity pattern. These spectral features are sensitive to the preprocessing. The power spectra density is usually taken as the oscillatory energy per frequency produced by the neural system. The cross spectra are the bivariate cross-covariance of the frequency information. Both the power spectra and the cross spectra are the second order statistical properties calculated with the artifact removed data.

As widely accepted by the MEEG community, the power spectra curves represent the neural oscillation, and the log transformed spectra allow for the better visualization of high frequency low amplitude oscillations. Most of the linear and nonlinear coupling measures can be derived as the functions of cross spectra, such as the partial/lagged coherence, phase lag index, phase locking, and power envelope correlation [21]. This may mean that the cross spectra contain all second order information of electrophysiology dynamics by completely determining the statistical properties.

The later stage of electrophysiology pipeline is to analyze the neural oscillatory behaviors from the band limited power spectra or to map the source connectivity by firstly solving the inverse problem with the cross spectra [22]. Finally, MEEG researchers typically use statistical learning methods to find the cognitive or pathological relevance from the oscillation parameters and the connectivity map. Thus, the quality assurance of cross spectra or the frequency information is a critical step in the large scale MEEG processing. However, the quality control of frequency spectra is unfortunately ignored.

## 1.6 The structure of this paper

In this paper, we initiate by presenting the Parallel Log Spectra (PaLOS) phenomenon, and introduce an indicator named PaLOSi, which employs conventional Principal Component Analysis (PCA) for quantifying this phenomenon. Subsequently, we validate PaLOSi on an electroencephalogram (EEG) database incorporating various factors. Thirdly, we proceed to simulate a multitude of scenarios potentially giving rise to the PaLOS phenomenon through dipole simulations. Then, we conducted testing to observe changes in PaLOSi values of the source activities in simulated and real EEG data with high PaLOSi values. Finally, we undertake a statistical analysis of the network weight distribution for EEG recordings with high PaLOSi value.

# 2. Results

## 2.1 The presence of PaLOS

### 2.1.1  The initial observation of PaLOS

The spectral profile characterizes the time invariant information which reflects the intrinsic dynamics of neural system at the macro level. Although it is impractical to control each preprocessing step, checking the spectra can give the overall evaluation of the retained brain signals as an alternative to indicate the global effect of artifact removal.

The initial observation of PaLOS was detected in an oversites EEG quality comparison study where two subjects from the Child Mind institute underwent the EEG recording in Midtown Manhattan and Staten Island, New York state. The resting EEG was recorded under the eyes-open/closed condition using the EGI 128 channel system with Cz reference and 500 Hz sampling rate. All the recordings were sent to Automagic toolbox for artifact removal and correction. The waveform of subject L under 'Rest Run01_20190322165' was shown in Fig. 1 and its multichannel power spectra were estimated using the default parameter of 'pop_spectopo' function built in the EEGLAB. **Fig. 1**(a) illustrated that the 105 channels of power spectra were asymptotically parallel to each other; the power spectra in the log scale were shifted up and down with constant across all frequencies, or say, the power spectra in the natural scale were scaled up or down by multiplying or dividing constant across all frequencies; the spectral topographies are nearly identical. **Fig. 1**(b) displays a segment of waveform that the power spectra were estimated from. By manual scrolling the window and visual inspection, the cleaned EEG by Automagic showed good looking waveforms.  The quality assessment by using the temporal metrics, the absolute power, the relative power, the alpha peak, and the microstate spatial topography suggested that data quality was good, useful, and acceptable.

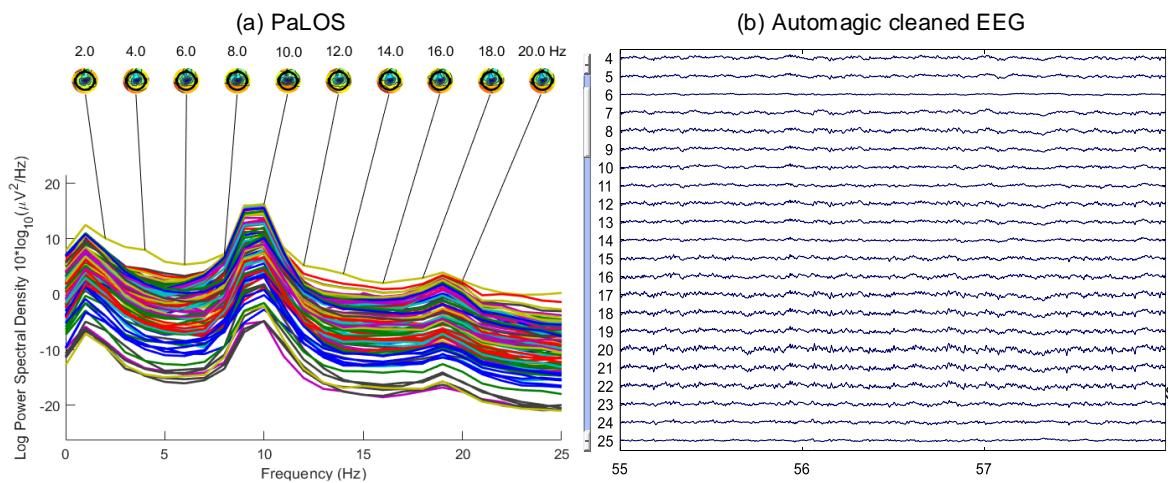

**Fig. 1 Illustration of PaLOS.** (a) 105 channels of power spectra at the log10 scale and the spectral topographies at the discrete frequencies with 2Hz interval; (b) display of EEG waveforms that were used for estimating the multichannel power spectra shown in (a). Note that the EEG waveforms were cleaned after running the Automagic pipeline.

### 2.1.2  Reproducible observations in multisite databases

One large size EEG database (HBN) and four small to medium-sized datasets (PDHC, MIPDB, DSTR, TRCS) underwent the Automagic preprocessing and received the quality labels of "Good", "Ok", and "Bad". The PaLOSi was computed for each preprocessed data from these five databases. **Fig. 2**A showed that the PaLOSi of preprocessed data primarily distributed in the range of [0.25, 0.65]; a significant portion of HBN EEG exhibited high PaLOSi, ranging from 0.9 to 1. The 8.9% of EEG exhibited the PaLOSi exceeding 0.7, with even 3.72 of EEG surpassing 0.9, although they were all labeled as "Good" by the temporal metrics in the Automagic pipeline (**Fig. 2**B). The data labeled as "Ok" and "Bad" but presenting the PaLOSi>0.7 accounted for 4.35% (**Fig. 2**C) and 7.61% (**Fig. 2**D) respectively. The 13.25 of EEG presented the PaLOSi>0.7 but were categorized as "Good" or "OK" cases. This revealed that the PaLOS seems to be commonly existed among the labeled "Good" and "Ok" cases; and the PaLOS phenomenon was reproducibly observed in the five databases. Thus, PaLOS was not an accidental case, demanding further investigation.

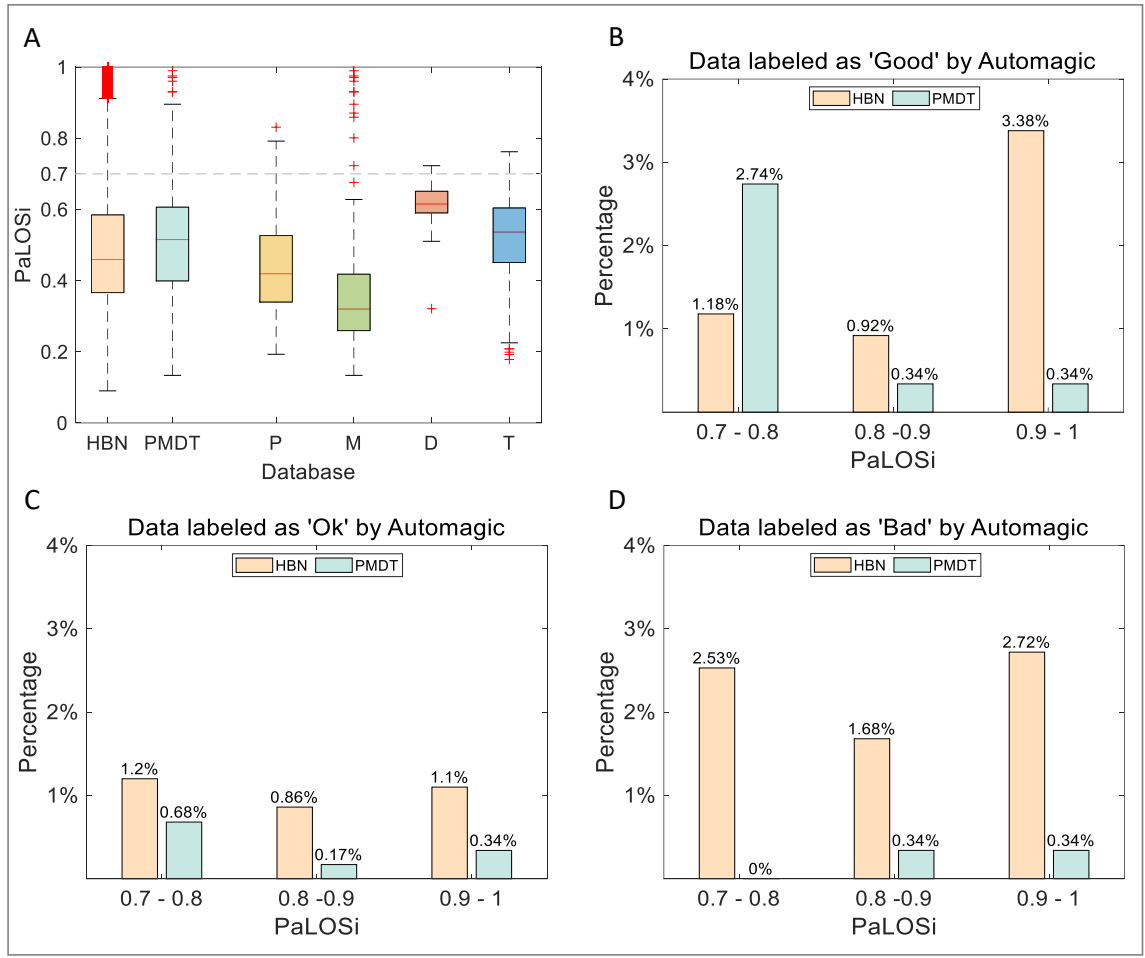

**Fig. 2 Reproducible observations of PaLOS in multisite databases.** *A. the PaLOSi values of preprocessed EEG data from the HBN (2951), the PDHC (41), the MIPDB (118), the DSTR (65), the TRCS (60); B, C, D: the percentage of PaLOSi values greater than 0.7 and labeled as 'Good', 'Ok', and 'Bad' data by Automagic, respectively. Note that PaLOS phenomenon was considered reproducibly observed if PaLOSi>0.7, and all the percentiles are the ratio of number of investigated cases to the total number of instances from the five databases. Due to the relatively small sample size of the databases except for the HBN, they were put together as the PMDT database for better visualization.*

## 2.2 What does PaLOS mean from the simulative perspective?

### 2.2.1 The loss of brain activities

Using real source activities, clean EEG (CE) was obtained through forward modeling. Subsequently, independent component analysis (ICA) was performed, removing some components to generate excessively preprocessed EEG (EPE) with PaLOSi values exceeding 0.7. Then, the sLORETA method was employed for inverse solution to obtain SCE and SEPE. PaLOSi values were computed for each stage. The Fig. 3B summarizes the results. We observed that the PaLOSi value of the ground-truth (GT) lies in the range of [0.1, 0.2], indicating strong heterogeneities in source activities. After forward calculation, the PaLOSi value increases, that is the PaLOSi value of CE ranging from 0.3 to 0.4. However, the reconstructed source activities in SCE exhibit a reduced PaLOSi value compared to CE (p < 0.001). Similarly, for EPE, the PaLOSi value of the corresponding source activities (SEPE) is lower than that of EPE (p < 0.001). It's noteworthy that although the source PaLOSi value decreases in comparison to the scalp PaLOSi, this reduction is minor.

Furthermore, Fig. 3A depicts the variation of PaLOSi values in the scalp and the source level as a function of the number of removed ICs. It is evident that PaLOSi values gradually increase with a greater number of removed ICs. This indicates that the loss of brain signals leads to an increase in PaLOSi values, with a higher degree of signal loss corresponding to higher PaLOSi values.

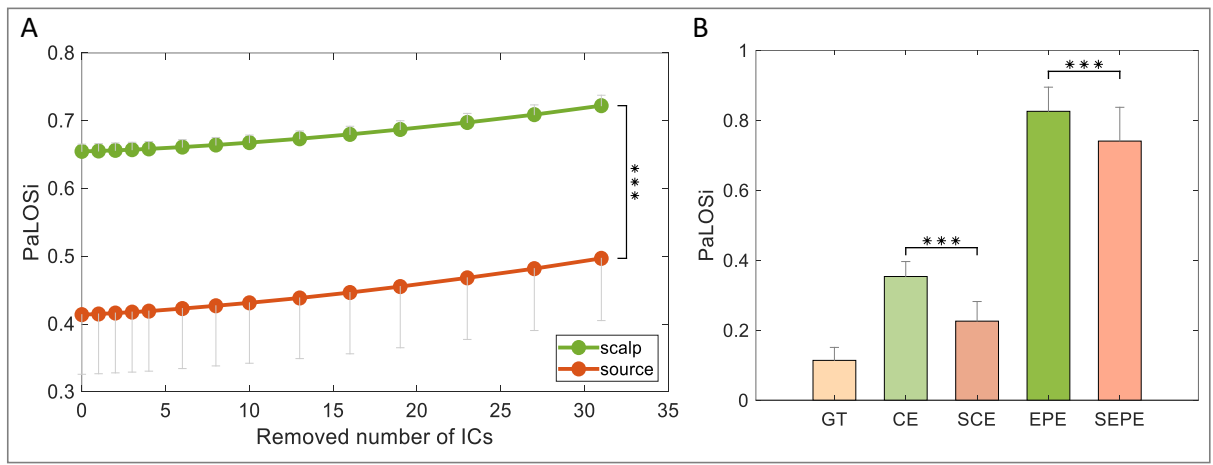

*Fig. 3 **The PaLOSi Values at different stages.** A: variation of scalp and source PaLOSi values with the number of removed ICs. The green line: scalp PaLOSi values, the red line: source PaLOSi values. C: the scalp and source PaLOSi values. GT: ground-truth of source, CE: clean EEG, SCE: source of CE, EPE: excessive preprocessed EEG, SEPE: source of EPE. \*\*\*: p < 0.001.*

### 2.2.2   The effect of volume conduction model

The volume conduction model is a biological signal propagation model that considers brain tissue, skull, and scalp as mediums, where the electrical signals on the scalp are a linear combination of activities from numerous brain sources. In Fig. 3B, the PaLOSi value of CE is significantly lower than that of EPE, and the PaLOSi value from GT to CE does not show a substantial increase. Additionally, the PaLOSi values from CE to SCE and EPE to SEPE exhibit only a modest decrease. These observations suggest that the potential cause of elevated PaLOSi values may be attributed to a significant loss of brain signals, with minimal discernible impact from volume conduction intervention.

### 2.2.3   The disruption of brain functional connections

Due to the PaLOSi quantifies the homogeneity of the spectrum, and most network connectivity methods are based on spectral characteristics, EEG data with high PaLOSi values may introduce distortions in network analysis. We constructed networks using coherence as the connectivity measure. On the source level, we compared the network similarity between GT and SCE, GT and SEPE, and SCE and SEPE. On the scalp level, we compared the network similarity between CE and EPE. The results are illustrated in Fig. 4A.

On the source level, the high network similarity, reaching 0.85, is observed between GT and SCE, while the similarities between GT and SEPE and between SCE and SEPE are low, with GT and SEPE exhibiting the lowest similarity. On the scalp level, CE and EPE have a similarity of approximately 0.67. It can be observed that network similarity (Fig. 4A) corresponds to the PaLOSi results (Fig. 3B). Smaller differences in PaLOSi result in higher network similarity, and conversely, greater differences in PaLOSi result in lower network similarity. This result suggests that EEG data with high PaLOSi values can lead to network distortions.

In Figure 4B, C, and D, the skewness statistics of the network weight distribution, the mean network weight statistics, and the PaLOSi value statistics for CE and EPE are presented. The pairwise Pearson correlations between these measures are also described. Figure 4B indicates that the average network weight distribution of CE falls between 0.2 and 0.3, with a positive skewness, suggesting a right-skewed distribution where the network weights are concentrated on the left side of the mean network weight. In contrast, the average network weight distribution of EPE ranges from 0.5 to 0.7, with a predominantly negative skewness, indicating a left-skewed distribution with network weights concentrated on the right side of the mean network weight.

Simultaneously, it is noteworthy that both for CE and EPE, there exists a significant negative correlation between the skewness of the network weight distribution and the mean network weight ($p < 0.001$). Additionally, it is worth mentioning that there is a negative correlation between skewness and PaLOSi values ($p < 0.001$), while a positive correlation exists between mean network weight and PaLOSi ($p < 0.001$). This implies that as PaLOSi values increase, the average weight of the network becomes larger, skewness decreases, indicating a prevailing presence of larger weights in the network.

Network connectivity methods are typically built upon spectral characteristics. Therefore, high PaLOSi values may lead to inaccuracies in network connectivity, resulting in the disruption of network connections, and subsequently impacting the credibility of subsequent brain network analysis and research findings.

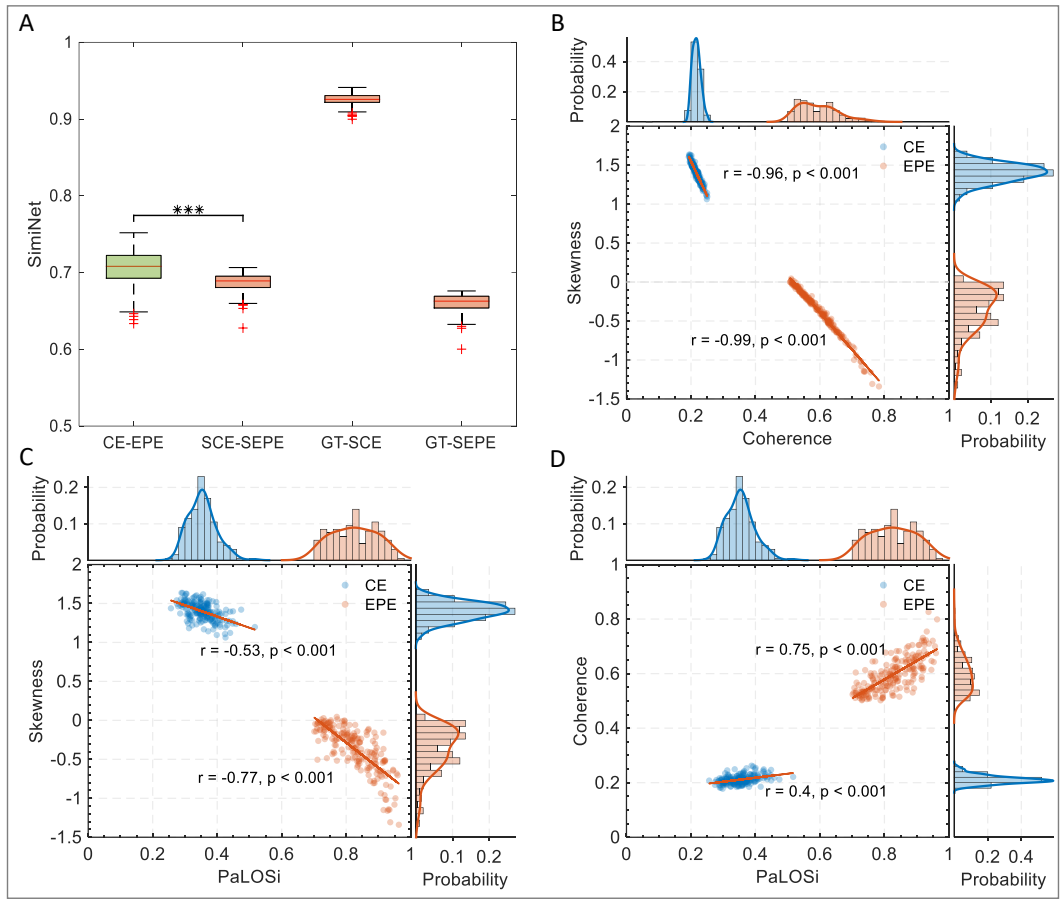

*Fig. 4* ***The similarities of brain networks, average network weight, the PaLOSi value and the skewness of network weight distribution for simulated data.*** *A: the similarities of brain networks at different stages. B, C, D: the Pearson correlation and distribution of skewness, average network weight, and PaLOSi. \*\*\*: p < 0.001.*

### 2.2.4    The dipole configuration with local dominance

In order to investigate the emergence of the PaLOS phenomenon, we conducted dipole simulations and Table 1 presents the detailed configuration of dipoles. Fig. 5 displays the PaLOS and non-PaLOS phenomenon. It is apparently seen from the panel A of the Fig. 5 that when merely a sole dipole is activated, the multichannel log power spectra are parallel, and the PaLOSi value is 1. Fig. 5B showed when there are three dipoles distributed in a dispersed manner, the trajectories of the multi-channel log power spectrum intertwine, and the PaLOS phenomenon does not exist, and the PaLOSi value is 0.6640.

To circumvent the influence of the number of dipoles, we placed 19 radial dipoles on the cortical surface, as illustrated in Fig. 5C and Fig. 5D. One dipole was positioned in the occipital lobe region as the center, with the remaining 18 dipoles organized into three groups of six each. These groups formed concentric circles from the center, with the first group having the smallest radius and the third group the largest. Fig. 5C depicts a scenario where dipoles have an intensity of 1 and coherent connections range from 0.8 to 0.9. In this situation, an approximation of the PaLOS phenomenon arises, with a source PaLOSi value of 0.9121 and a scalp PaLOSi value of 0.9899. This indicates that coherent connections between sources can lead to parallel power spectra among sources, even extending to the scalp. In Fig. 5D, the dipole distribution is the same as in Fig. 5C, but with a notable difference. First, the dipoles diffuse outwards from the central dipole.

Additionally, the dipole strengths exhibit Gaussian decay, with the central dipole having a strength of 1, the nearest group of six dipoles having a strength of 0.98, the second group having a strength of 0.05, and the third group having a strength of 0.01. The coherent connections between dipoles range from 0 to 0.1. Table 1 reveals a source PaLOSi value of 0.1646 and a scalp PaLOSi value of 0.9039 for this scenario. An approximation of the PaLOS phenomenon emerges on the scalp. This suggests that even if the PaLOS phenomenon is not evident at the source level, the aggregation of strong dipoles can lead to overwhelming dominance and, consequently, the emergence of the PaLOS phenomenon on the scalp.

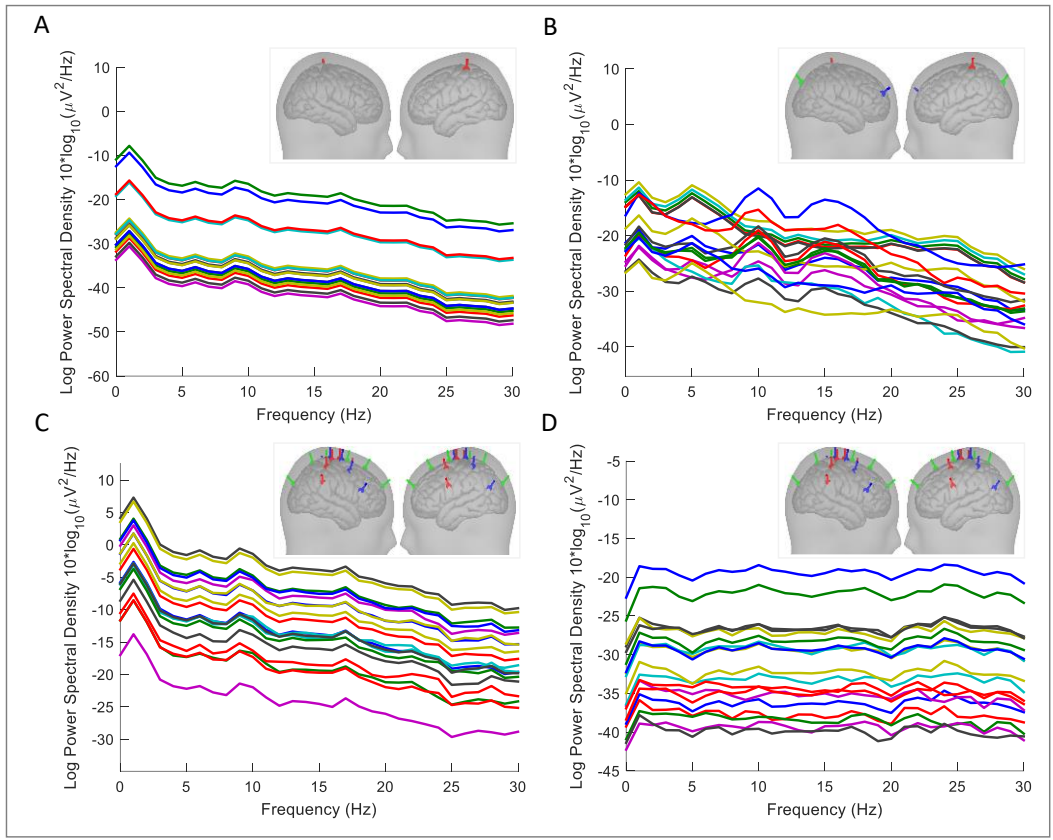

Fig. 5 **Various simulated scenarios with the PaLOS and non-PaLOS phenomenon.** *Each panel presents the trajectory of the log power spectrum, accompanied by views of the brain from both the left and right sides in the upper-right corner. The cortical surface displays the positions and radial orientations of dipoles.*

*Table 1. The detailed dipole configuration information along with scalp and source PaLOSi values.*

| Panel Configurations | A | B | C | D |
|---|---|---|---|---|
| Number of dipoles | 1 | 3 | 19 | 19 |
| Orientation of dipoles | | | Radial | |
| Distribution of dipoles | ~ | Dispersion | Concentric circular | Concentric circular |
| Intensity of dipoles | 1 | 1 | 1 | Gaussian decay |
| Connectivity values of dipoles | ~ | $0-0.1$ | $0.8-0.9$ | $0.1-0.2$ |
| Dimension of the leadfield | 19 X 1 | 19 X 3 | 19 X 19 | 19 X 19 |
| Is the spectrum parallel? | Yes | No | Yes | Yes |
| Source PaLOSi | ~ | ~ | 0.9121 | 0.1646 |
| Scalp PaLOSi | 1 | 0.6640 | 0.9899 | 0.9039 |

## 2.3 Evidence from the real instances

### 2.3.1 The distribution of the normative coherences and relationship with PaLOSi

The results of PaLOSi values, average network weights, and the skewness of network weight distribution are reported for a total of 1966 cases from multiple countries and sites, as depicted in Figure 6. Across various frequency bands, the skewness is generally greater than 0, average network weights are distributed approximately between 0.1 and 0.4, and PaLOSi values are distributed around 0.3 to 0.7. Furthermore, there is a negative correlation between skewness and average network weights, as well as between skewness and PaLOSi ($p < 0.001$). Conversely, there is a positive correlation between average network weights and PaLOSi ($p < 0.001$). These findings align with the results observed in the CE data simulations.

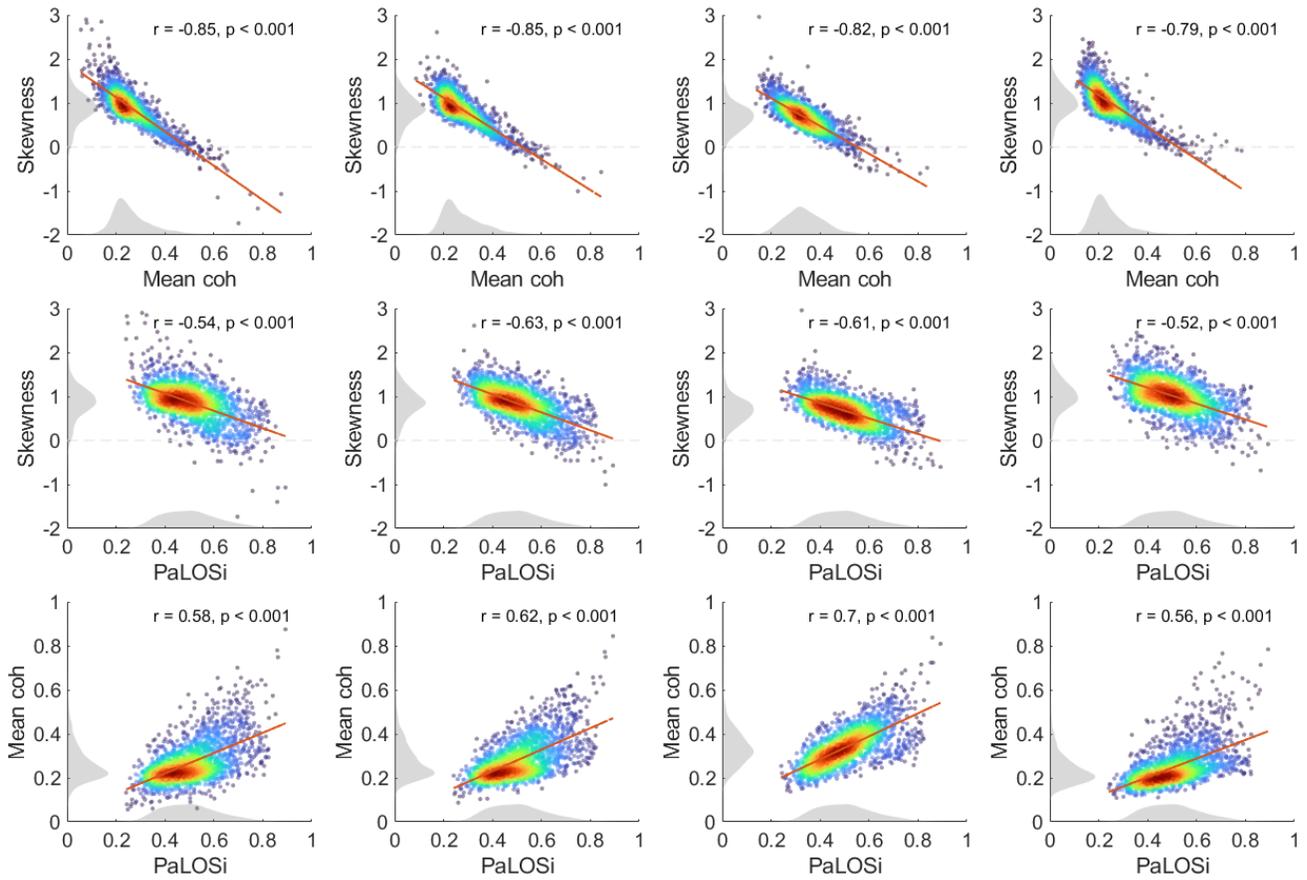

*Fig. 6* **PaLOSi values and the network weight distribution at each frequency band of normative data.** *From left to right are the δ, θ, α, and β frequency bands. The brightness in each subplot represents the degree of data aggregation, and the grayscale image on the coordinate axes represents the distribution of the corresponding indicators.*

### 2.3.2 The coherence distribution of data with high PaLOSi values

Figure 7 illustrates the distribution of network skewness, average network weight, and PaLOSi values, as well as the pairwise correlations among them, for data with PaLOSi values greater than 0.7 from HBN and PMDT databases processed through Automagic preprocessing. It can be observed that the average network weights are relatively large, predominantly distributed between 0.6 and 1. The skewness, for the most part, is negative, indicating a left-skewed network where weights are concentrated in regions greater than the mean. Similarly, there is a significant negative correlation between skewness and both average network weight and PaLOSi, while there is a significant positive correlation between average network weight and PaLOSi.

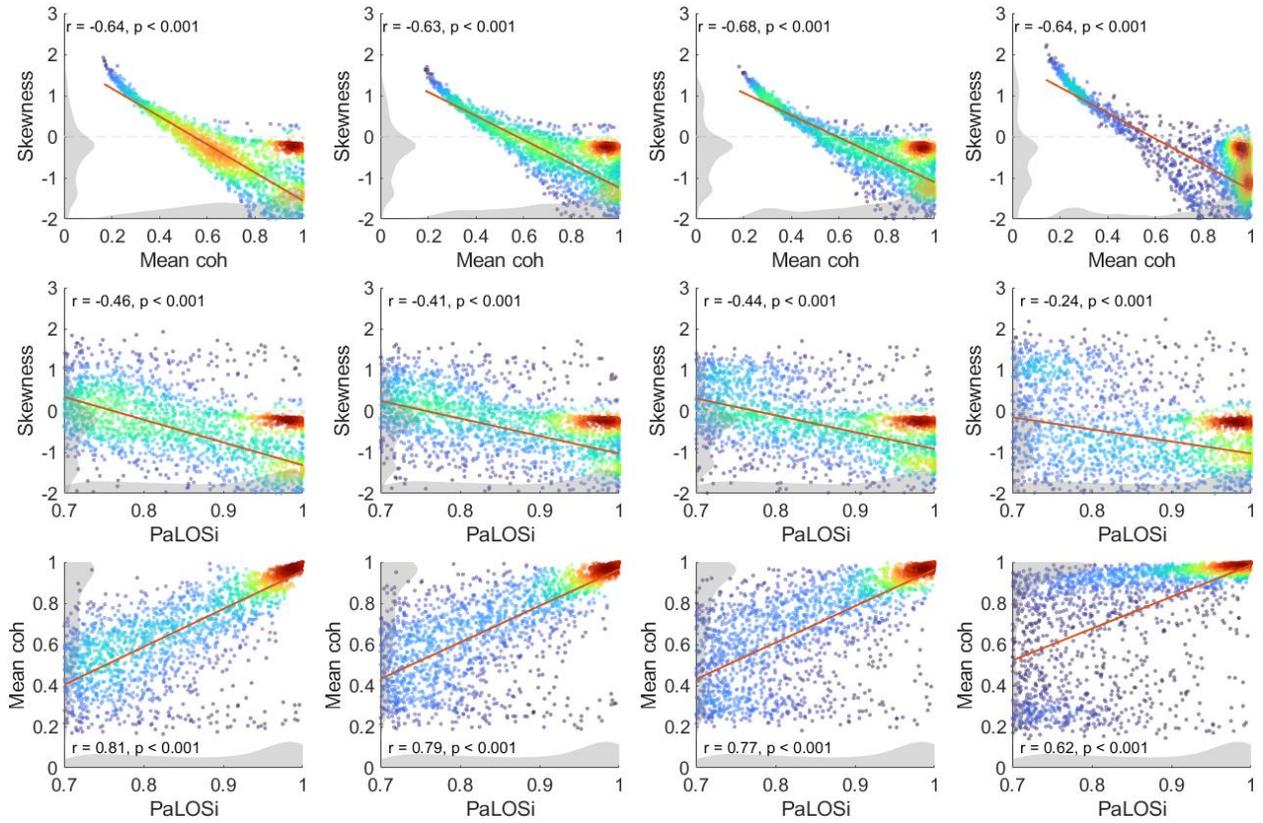

*Fig. 7 **The network skewness, average network weight, and PaLOSi values of HBN and PMDT data with high PaLOSi values (greater than 0.7), processed through Automagic preprocessing.** From left to right are the δ, θ, α, and β frequency bands. The brightness in each subplot represents the degree of data aggregation, and the grayscale image on the coordinate axes represents the distribution of the corresponding indicators.*

### 2.3.3 The spectral loss submerged in the EEG waves

High PaLOSi values have been detected in preprocessed EEG data from multiple databases. Here, we have selected one case from each database for analysis. Fig. 8 and Table 2 respectively present individual cases from five different databases and provide detailed information.

*Table 2. Detailed information of individual cases along with their scalp and source PaLOSi values.*

| Figure Information | 8A | 8B | 8C | 8D | 8E | 8F |
|---|---|---|---|---|---|---|
| Database | HBN | HBN | PDHC | MIPDB | DSTR | TRCS |
| Subject | NDARHN078CDT | NDARHR386KRJ | Control1291 | A00054597 | Sub-086 | Sub-46 (session 3) |
| Age | 8.53 | 21.5 | 60 - 79 | 6 - 9 | 20 | 21 |
| Gender | F | F | Unknown | F | F | M |
| State | Eye-open | Eye-open | Eye-open | Resting | Eye-close | Eye-open |
| Label | 'Good' | 'Good' | 'Ok' | 'Good' | 'Good' | 'Good' |
| Scalp PaLOSi | 0.9998 | 0.9985 | 0.7920 | 0.9737 | 0.7191 | 0.7080 |
| Source PaLOSi | 0.8792 | 0.8756 | 0.4590 | 0.5309 | 0.5655 | 0.6057 |

Fig. 8A and Fig. 8B showcase individual cases from the HBN database, characterized by a dense power spectral density curve. Moreover, the EEG waveforms across all channels appear nearly identical, suggesting that the signals are consistent across all channels at each time point. The scalp PaLOSi values are exceptionally high, approaching 1. The case depicted in Fig. 8C is from the PDHC database, with multi-channel power spectral trajectories displaying a pronounced dominance around 2Hz. The scalp PaLOSi value is 0.7920, and some channels still exhibit similar waveforms, as illustrated in the background image. In Fig. 8D, the individual case is from the MIPDB database, with power spectral density demonstrating a parallel phenomenon. The PaLOSi value is 0.9737, similar to the HBN cases, where the EEG waveforms across various channels are nearly identical. Cases in Fig. 8E and Fig. 8F are from the DSTR and TRCS databases, respectively. In both cases, multi-channel power spectral trajectories exhibit a distinct dominance around 10Hz. The displayed EEG waveforms appear relatively normal, yet their scalp PaLOSi values are 0.7191 and 0.7081, respectively.

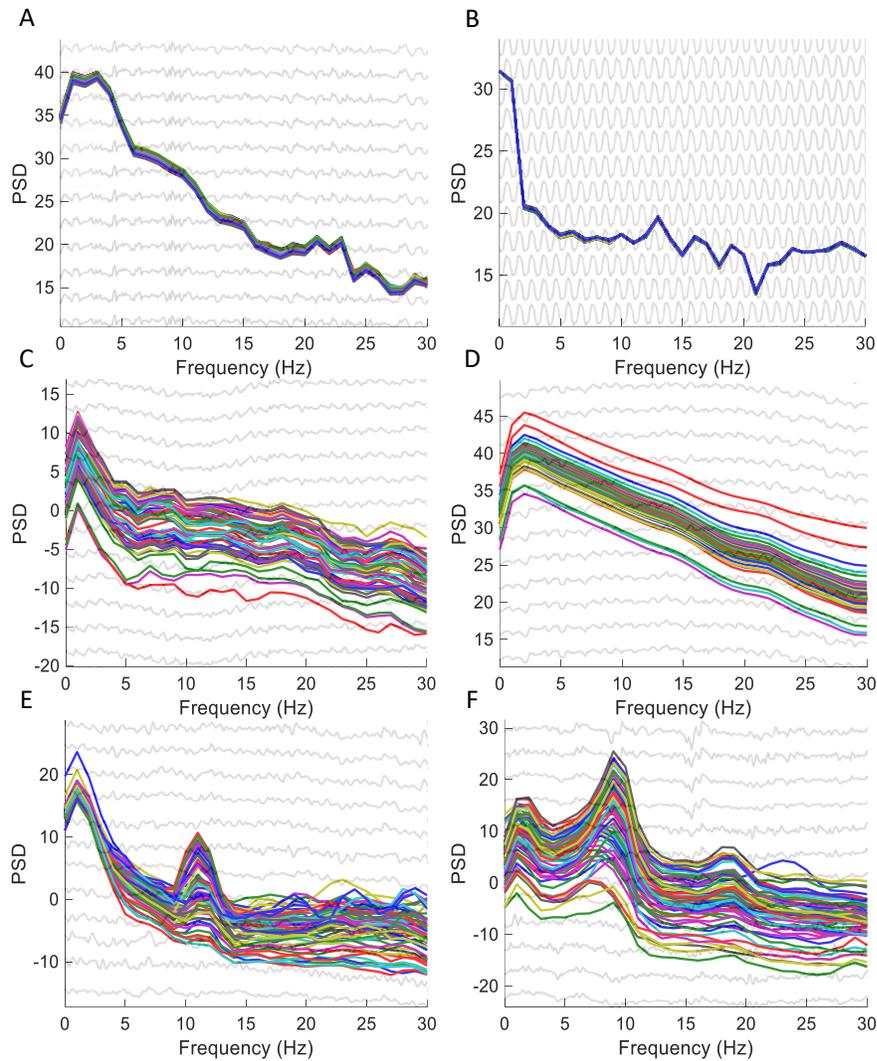

*Fig. 8 **Individual cases from 5 databases, where A and B are from the HBN database, and C to F are from the PMDT database.** Each panel presents the power spectral trajectory along frequencies and its EEG waveform of a segment (gray background). PSD: power spectral density.*

## 3. Discussion

In this study, we firstly introduced the issue of Parallel Log Spectra (PaLOS) and proposed a metric, PaLOSi, to quantify it. Subsequently, we validated the PaLOS phenomenon in a dataset comprising 30,094 samples from multiple databases and in the normative database, indicating that PaLOS is not an isolated occurrence by accident. Thirdly, we employed dipole simulations to explore various scenarios that could lead to PaLOS and examined the source PaLOSi of the EEG exhibiting high PaLOSi values. Next, we investigated the impact of the EEG with high PaLOSi on its brain network functional connections. Finally, we analyzed individual cases from multiple databases, including their power spectra and EEG waveforms.

Subsequently, both simulated and real data indicated that EEGs with higher PaLOSi values exhibit network weights concentrated in regions with larger weights. We acknowledge the unreliability of this phenomenon. Although the standards for cleaning EEGs are currently unclear [23], and there is a lack of baseline real data for comparison, we sought to determine the differences in network weight distribution between clean EEGs and EEGs with high PaLOSi values through simulated data and normative data compared with HBN and PMDT data results. Furthermore, we explored the correlations between the skewness of network weight distribution and mean network weight, skewness and PaLOSi values, as well as the correlation between mean network weight and PaLOSi (Fig. 4, Fig. 6 and Fig. 7). In coherence-based network construction method, PaLOSi showed a significant negative correlation with skewness and a significant positive correlation with mean network weight.

It is well known that the acquisition of electroencephalogram (EEG) data is susceptible to various sources of noise, significantly distorting the recorded signals [24]. Therefore, ensuring the acquisition of clean EEG data has persistently remained a challenging task. Despite the current availability of several preprocessing pipelines, such as the Harvard EEG Automated Processing Pipeline (HAPPE) [12], the Batch EEG Automated Processing Platform (BEAPP) [13], the automatic pre-processing pipeline (APP) [11], PREP (Bigdely-Shamlo et al. 2015), Automagic [14], the EEG Integrated Platform Lossless pre-processing pipeline (EEG-IP-L) [17], the pursuit of clean EEG data remains an unresolved objective in the field. Given this context, contemplating how to remove noise to the greatest extent possible while preserving brain signals to the maximum degree is a question worthy of deep consideration [17], and careful preprocessing is self-evident.

The PaLOS phenomenon proposed in this paper is undeniably present in preprocessed EEG data, and the PaLOSi index can detect this phenomenon (Fig. 2). Unlike many time-domain quality metrics such as signal amplitude, kurtosis, and max gradient [23,26,27], PaLOSi focuses on the spectral perspective to examine the quality of multi-channel power spectra. Obviously, while some EEG data may pass certain time-domain quality metrics and receive labels like 'Good' or 'Ok,' the PaLOSi index provides a unique perspective, leaving no room for such data to escape detection (Fig. 8).

In order to understand the parallelism of power spectra in source activity corresponding to high PaLOSi EEG, we conducted tests on both simulated and real data. The PaLOSi metric quantifies the homogeneity or heterogeneity of the cross-spectra in the frequency scale. A larger PaLOSi indicates greater homogeneity, resembling parallel power spectra, while a smaller PaLOSi suggests more suppression in the spectra. Fig. 3A and Fig. 3B suggest that the loss of brain signals may lead to the spectra becoming more homogeneous, resulting in an increase in PaLOSi values. Additionally, from Fig. 3B, it is evident that the spectra of source activities are more heterogeneous compared to scalp EEG. Furthermore, the increment of PaLOSi from source to CE may suggest that the volume conduction effect may introduce some homogeneity information to the spectrum. After applying ICA to CE and discarding some ICs to obtain EPE, the PaLOSi values increase. This suggests that there might be some association between the PaLOSi derived from PCA and ICA. To explore this further, we simulated 200 CE datasets, then applied ICA to these datasets, ranked ICs based on explained variance, and examined how scalp and source PaLOSi values change as ICs are removed. The results are presented in Fig. 3A. It can be observed that the PaLOSi values slowly increase as ICs are removed, indicating a gradual linear relationship between PCA and ICA. Importantly, the source PaLOSi values consistently remain lower than the scalp PaLOSi values, which further confirms the findings in Fig. 3A.

We explored the source configurations that could lead to the PaLOS phenomenon, as shown in Fig. 5. It is evident that, when comparing Fig. 5A and Fig. 5B, the PaLOS phenomenon may result from either a single source or multiple sources clustered together. However, such configurations are not common from a physiological perspective. Subsequently, we simulated scenarios with multiple dispersed sources, some of which may cluster together, as illustrated in Fig. 5C and Fig. 5D. On one hand, when certain sources cluster together while others are dispersed across different cortical regions, and the coherence between each pair of sources ranges from 0.8 to 0.9, the source PaLOSi reaches 0.9121, and the scalp PaLOSi value is 0.9899, indicating the emergence of a phenomenon resembling parallel power spectra (Fig. 5C). On the other hand, when some sources cluster together while others are dispersed across different cortical regions, and the source strengths follow a Gaussian decay pattern from the center towards the periphery, the source PaLOSi is 0.1646, and the scalp PaLOSi value is 0.9039. This also exhibits a phenomenon similar to parallel power spectra (Fig. 5D).

Fig. 8 presents individual cases from multiple databases. Firstly, cases from the HBN and MIPDB databases exhibit phenomena resembling parallel power spectra. Across various frequencies, the power spectral density varies very little, resembling a thick curve. Furthermore, the EEG waveforms for most channels are nearly identical at different time points. This might be due to a scenario where there is only one source or multiple sources clustered together (as shown in Fig. 5A) or potentially a situation with highly coherent connections between sources (as shown in Fig. 5C). On the other hand, the cases from the DSTR and TRCS databases are more likely to correspond to the scenario depicted in Fig. 5D. This seems to align better with physiological background. The case from the PDHC database may be a combination of the scenarios in Fig. 5C and Fig. 5D, as some channel EEG waveforms are similar, while others are not. Additionally, we observed that the cases from HBN and MIPDB databases have much higher PaLOSi values compared to the other cases. It's noteworthy that the cases from HBN and MIPDB databases are in the age range of 6-9 years, indicating that EEG recordings from children may be more prone to PaLOS issues. When collecting EEG data from children, it is important to pay close attention to this phenomenon.

We do not deny the possibility of other factors contributing to PaLOS issues. However, it is evident that EEG data with high PaLOSi values can introduce unreliable networks, significantly impacting subsequent analyses. The PaLOSi metric we introduced serves as a valuable tool to address this issue effectively.

# 4. Methods

## 4.1 large sample of resting EEG

### 4.1.1    Oversite EEG recording for quality comparison

The human subjects A and L from the Child Mind Institute took the resting EEG recording with Brain Products (BP), Electrical Geodesics Inc. (EGI) in the two sites of Midtown Manhattan where the Child Mind Institute is located, and the Staten Island. The BP system was with 64 channels and 5000 Hz sampling rate, while the EGI system was with 128 channels and 500 Hz sampling rate. The recording for each subject comprised of multiple sessions and runs. All the raw recordings were cleaned and corrected by the Automagic pipeline that is detailed in the section 4.2. The quality of cleaned EEG recordings was comprehensively assessed by the combination of the temporal metrics, the absolute power, the relative power, the alpha peak, and the microstate topographies.

### 4.1.2    The HBN EEG

The Healthy Brain Network (HBN) biobank [1] accrued by child mind institute released the raw resting EEG recording of thousands of children and adolescents aged 5-21 residing in the New York city. The resting EEG of 2961 subjects from the early releases of HBN were analyzed in this study. The resting EEG was recorded with Cz reference, 500Hz sampling rate using 128-channel system by Electrical Geodesics Inc. The public data per subject contains the raw recorded EEG and preprocessed EEG by Automagic pipeline.

### 4.1.3    The PMDT EEG

- **P**: The PDHC (Parkinson's Disease and Healthy Control) [28] EEG database encompasses a total of 41 PD patients and 41 healthy controls from the University of New Mexico, as well as the University of Iowa. In this study, healthy controls were utilized. Resting state EEG data from subjects aged between 60 and 79 were acquired with eyes open, utilizing a 64-channel Brain Vision system at a sampling rate of 500 Hz.

- **M**: The Multimodal Resource for Studying Information Processing in the Developing Brain (MIPDB) EEG dataset[29] provided the electrophysiology dataset and the eye-tracking dataset by the Child Psychology Institute. Participants were from the New York metropolitan area, aged 6-44 years, using a 128-channel EEG Geodesic Hydrocel system with a sampling rate of 500 Hz and the online recording reference at the Cz.

- **D**: The Pupillometry and Electroencephalography in the Digit Span Task and Rest (DSTR) EEG dataset (Pavlov et al., 2022) comprised 88 subjects, 23 out of which had no resting EEG. Each subject underwent 4 minutes of EEG recordings under the resting-state and eyes closed condition, at a sampling rate of 1000 Hz, using 64 channels based 10-10 system and the reference electrode at FCz. The subjects spanned in the age from 17 to 25.

- **T**: The Test-Retest Resting and Cognitive State (TRCS) EEG dataset (Wang et al., 2022) was compiled by a team from the Department of Psychology at Southwestern University in China. They collected EEG data from 60 participants three times, each time comprised of 1 eye-open recording and 1 eye-closed recording, conducted at an interval. Thus, a total of 360 resting-state EEG recordings were collected. The age of all participants ranged from 18 to 28, and the recordings were obtained using 64 channels based on the 10-20 international standard electrode placement system.

The sampling rate was 500 Hz, and FCz served as the reference electrode. In the resting-state EEG recordings, both the resting open-eye state and closed-eye state were recorded for 5 minutes each from all subjects. Subsequently, to unify the channels across various electrode cap sizes, the channel count was reconfigured to 62.

### 4.1.4 The normative EEG database

The normative database is derived from a multicenter qEEG normative project [30], with data collected from 1966 participants across 9 countries and 14 sites. The data recordings utilized 19 channels based on the 10/20 International Electrode Placement System: Fp1, Fp2, F3, F4, C3, C4, P3, P4, O1, O2, F7, F8, T3/T7, T4/T8, T5/P7, T6/P8, Fz, Cz, and Pz. The frequency range covered is from 1.17 to 19.14Hz with a resolution of 0.39Hz. The cross-power spectra of scalp EEG were calculated using Bartlett's method by averaging the periodograms of more than 20 consecutive non-overlapping segments. Each site conducted preprocessing on the data, excluding Independent Component Analysis (ICA). The shared dataset includes mutual power spectral data of EEG, as well as anonymized participant information, such as age and gender, along with technical parameters like EEG recording equipment, electrode placements, EEG recording references, laboratory details, and the country of data collection.

## 4.2 Automatic preprocessing

Automagic [14] is an open-source, large-scale automated preprocessing MATLAB software platform with a user-friendly interface that allows users to preprocess EEG data by configuring a series of preprocessing parameters. The Automagic pipeline comprises three parts: EEG data import and parameter configuration, EEG preprocessing, and EEG quality assessment. In the preprocessing phase, bad channel detection (e.g., PREP pipeline, clean_rawdata(), or identifying channels with standard deviation exceeding a set threshold), filtering, artifact correction (e.g., EOG regression, multiple artifact rejection algorithms, or robust PCA-based noise removal), and DC offset removal are included. Subsequently, quality assessment can be performed after the interpolation of the preprocessed data.

## 4.3 Quality control metrics

### 4.3.1 Temporal aspect measures

In the case of Automagic, it provides four quality measures: the ratio of bad channels (RBC), the ratio of data with overall high amplitude (OHA), the ratio of timepoints of high variance (THV), and the ratio of channels of high variance (CHV). RBC indicates the ratio of identified bad channels to interpolated channels, with the number of interpolated channels being proportional to EEG quality. OHA quantifies the ratio of data points with high absolute voltage amplitudes. THV reflects the time point ratio of the standard deviation of all channels with higher voltage values. CHV indicates the ratio of channels with higher standard deviation of voltage at all time points. By selecting quality assessment metrics (RBC, OHA, THA, CHV) and specifying thresholds, the preprocessed data are automatically labeled as 'Good', 'Ok', or 'Bad'. In short, these metrics assess signal quality in the time domain, specifically considering that the voltage amplitude of the brain signal should fall within a certain range.

In this study, we preprocessed the aforementioned database using the Automagic toolbox, which involved bad channel detection, bad channel removal, filtering, and artifact correction. Then, any channels that were detected as bad channels were interpolated, and quality assessment was conducted. Preprocessing step parameters were set to those recommended by Automagic, and quality assessment thresholds were maintained at their default values (OHA: Good < 0.1 < Ok < 0.2 < Bad, THA: Good < 0.1 < Ok <0.2 < Bad, CHV: Good < 0.15 < Ok < 0.3 < Bad, RBC: Good < 0.15 < Ok < 0.3 <Bad). In this study, we calculated PaLOSi values for all preprocessed data and counted the proportion of data instances with PaLOSi values exceeding 0.7.

### 4.3.2 PaLOS index as the 1st frequency metric

The cross-spectral matrix is theoretically the sample covariance of multichannel Fourier series [31]. The multichannel Fourier series subscripting by the frequency $\omega$ is expressed as $\mathbf{\Phi}_\omega \in \mathbb{C}^{N_e \times N_s}$, with $N_e$ and $N_s$ denoting the number of electrodes and the number of segments, respectively. Assuming the spatial whitening matrix is $\mathbf{\Gamma}^\dagger$ and the diagonal matrix is $\mathbf{\Lambda}_\omega$, then the principal component decomposition on the multichannel Fourier series is

$$\mathbf{\Psi}_\omega = \mathbf{\Gamma}_\omega^\dagger \mathbf{\Phi}_\omega \tag{1}$$

$$\mathbf{\Gamma}_\omega^\ast \mathbf{\Phi}_\omega \mathbf{\Phi}_\omega \mathbf{\Gamma}_\omega = \mathbf{\Lambda}_\omega \tag{2}$$

$$\mathbf{S}_\omega \propto \mathbf{\Phi}_\omega \mathbf{\Phi}_\omega^\ast = \mathbf{\Gamma} \mathbf{\Lambda}_\omega \mathbf{\Gamma} \tag{3}$$

The presence of the cross-spectral homogeneity across frequencies can be attributed by both the prominent dominance of the largest eigenvalue and the frequency-invariant property of the eigenvectors. The largest eigenvalue corresponds to the largest explained variance in the principal component space. The whitening matrix is comprised of eigenvectors and identical to the principal decomposition at all the frequencies. This ensures that the cross spectral matrix at each frequency can be mapped into the common orthogonal coordinate spaces, helping disclose the cross spectral homogeneity. The spatial transformation applied to the Fourier series is identical across frequencies. The stepwise common principal component (CPC) decomposition method [32,33] can reduce the dimensions and sort the eigenvectors according to the explained variance simultaneously. The stepwise CPC analysis is applied to the cross spectra as

$$\mathbf{S}_\omega = \mathbf{\Gamma} \mathbf{D}_\omega \mathbf{\Gamma}^\dagger \tag{4}$$

and output the matrix $\mathbf{\Gamma}$ and the diagonal matrix $\mathbf{D}_\omega$. The PaLOSi is defined as

$$PaLOSi = \sum_\omega \max\{diag(\mathbf{D}_\omega)\} \Big/ \sum_\omega tr(\mathbf{S}_\omega) \tag{5}$$

Here, $\max\{diag(\cdot)\}$ is to extract the largest entry in the main dignoals, and $tr(\cdot)$ is the matrix trace operator. Thus, the PaLOSi lays in the range of [0, 1]. The larger the PaLOSi is, the more homogeneous the cross spectral matrices are across the frequencies. As an indicator of cross spectra, the PaLOSi can be applied to the source reconstructed activities as well as the scalp EEG activities.

### 4.3.3  Hybrid use of temporal frequency metrics

Totally 30,094 instances of resting EEG from the five databases were imported into the Automagic for preprocessing and each instance was eventually labeled as 'Bad,' 'Ok,' or 'Good'. The scalp EEG PaLOSi were computed for each instance. Then, the instances marked as 'Ok' or 'Good' by Automagic but with PaLOSi larger than 0.7 were selected for further investigation of their functional connectivity matrices.

### 4.4 Simulation

To understand what the PaLOS phenomenon indicates, the simulation and inference was implemented from two directions. The first was using the distributed source simulation to test whether the high PaLOS results from the loss of brain activities during the excessively preprocessing. Along this direction, the effects of preprocessing, volume conduction, source imaging on the PaLOSi and the functional connectivity matrices can be analyzed. The second direction was using the dipole fitting simulation to analyze the possible dipole configurations underneath the EEG waves that had high PaLOSi values.

### 4.4.1  The loss of brain activity in excessive preprocessing

The MNI iEEG atlas [34] contains eyes closed wakefulness resting potentials of accumulated 1772 intracranial electrodes at the healthy region from 106 subjects undergone surgery which nearly covered all the brain regions. Simulations were performed using real intracranial electroencephalography (iEEG) as the ground truth for the sources. The head model estimation employed the boundary element method, with a 64-channel electrode configuration covering the entire brain. The surface-based source space comprised 15,002 dipoles aligned with all cortical vertices. For computational ease, the cortical surface with 15,002 vertices was down sampled to 3002 vertices. Subsequently, 100 signal sources were randomly selected from intracranial EEG (iEEG) containing 1772 electrodes, and the remaining 2092 sources were simulated as low-intensity Gaussian noise. These 3002 signal sources were then randomly shuffled, ensuring that all dipoles on the cortex had source activity. Disregarding sensor noise, a noise-free electroencephalogram (CE) was synthesized through the volume conduction model. Finally, a CE was synthesized with the dimensions of 64 channels and 10,000-time samples. For statistical analysis, the same simulation procedure was repeated 200 times, with changes in source activity at each iteration.

To mimic the preprocessing, the independent component analysis (ICA) was applied to the CE. All the ICs were labeled as brain ICs and the non-brain ICs such as the channel noise, the muscle activity, etc. The ICs were sorted from smallest to largest according to the percentage of variance accounted for (PVAF) the CE. Then, the selected components were removed in combination while retaining the others explaining the first few highest variance. The back-projected EEG was taken as the excessively preprocessed EEG (EPE). For both CE and EPE, the source activities were obtained using the sLORETA method [35]. On the scalp level, we employed CE as a reference to compute PaLOSi values for both CE and EPE. On the cortex level, utilizing source activities as the GT, we computed PaLOSi values for GT, the source activities of CE (SCE), and the source activities of EPE (SEPE). For both scalp and cortex level, the functional connectivity matrices were estimated by using the coherence measure.

### 4.4.2 Dipole configurations

We used the MNI ICBM152 template anatomy and the aligned 19 electrodes with 10-20 system and estimated the head model by the boundary element method. The surface-based source space consists of 15002 dipoles aligned to all cortical vertices. To mimic the real source activity, we took the intracranial EEG potentials of iEEG atlas as the dipolar activity. The EEG potentials were generated by the forward calculation without noise considered. Subsequently, we employed the 'ft_dipolesimulation' function from FieldTrip [36] to perform dipole simulations, simulating cases with 1, 3, and 19 dipoles each, where the dipole orientations were set radially and the dipole strength was defined as the Euclidean norm. The specific configurations are as follows:

i.   1 single dipole with a radial orientation and a strength of 1.

ii.  3 dipoles with radial orientations, a strength of 1 each, and widely dispersed positions. The low coherence values were to make the 3 dipoles heterogeneous.

iii. 19 dipoles with radial orientations and a strength of 1 each. Centered around one dipole, the remaining 18 dipoles are arranged in three concentric circles radiating outward in groups of six. The coherence values between dipole activities were set to range from 0.8 to 0.9. The same strength and the high coherences of the 19 dipoles were to ensure that the dipole activities were homogeneous.

iv.  19 dipoles with radial orientations. With one dipole at the center, the remaining eighteen dipoles are arranged in groups of six, forming concentric circles that outward. Dipole strengths follow Gaussian decay from the center to the periphery, with the central dipole having a strength of 1. The coherence values between dipole activities were set to range from 0.1 to 0.2. The decayed strength and the low coherences were to ensure that the central dipole was prominently dominant among the 19 dipoles and all the dipole activities were heterogeneous.

## 4.5 Source reconstruction

The head model was constructed using the boundary element method, and the surface-based source space comprised 15,002 dipoles aligned with all cortical vertices. Source reconstruction was performed using the sLORETA method [35], followed by mapping the source activities to corresponding 79 ROIs using the Broadmann parcellation template in Brainstorm [37]. Ultimately, the principal component analysis (PCA) was employed for dimensionality reduction, resulting in source activity with a dimensionality of ROIs x time instances. The source activities associated with the ROIs enable the computation of PaLOSi values in the source space.

## 4.6 Functional connectivity analysis

The PaLOS phenomenon presented in the EEG data is unrealistic and raises concerns about data quality which may have impacts on the functional brain network analysis. The functional connectivity matrices were constructed using the coherence measure [38] in the frequency bands of delta (δ, 1–4 Hz), theta (θ, 4–8 Hz), alpha (α, 8–13 Hz), and beta (β, 13–30 Hz). Here, we conducted connectivity analysis on the simulated EEG datasets by 200 repetitions and real EEG datasets. Firstly, we identified instances from the HBN and PMDT databases that were marked as "Ok" or "Good" by Automagic and had PaLOSi values exceeding 0.7. Subsequently, we calculated the average network weight and skewness of the network weight distribution for simulated data, normative data, and the selected HBN and PMDT data instances. Furthermore, for the simulated dataset with known clean EEGs, we applied SimiNet [39] to measure the similarity between functional connectivity matrices under various conditions. We assessed the similarity of functional connectivity matrices between CE and EPE, SCE and SEPE, GT and SCE, GT and SEPE.

### 4.6.1 Skewness of the data distribution

Skewness is a measure of the degree of asymmetry in the statistical distribution of data, providing an indication of how the probability density curve deviates from symmetry relative to the mean. A positive skewness signifies that the shape of the data distribution has a longer tail on the right side of the mean, with the majority of the data concentrated on the left side of the mean. Conversely, a negative skewness indicates a longer tail on the left side, with the data concentrated on the right side of the mean. The skewness of a Gaussian distribution is zero. The skewness of the data distribution is defined as follows,

$$s = \frac{E(x - \mu)^3}{\sigma^3} \tag{6}$$

where $\mu$ is the mean of $x$, $\sigma$ is the standard deviation of $x$, and E represents the expected value. We computed the network of real data, calculated its average weight, and assessed the skewness of the network weight distribution. Subsequently, we investigated the correlation between average weight, skewness, and PaLOSi.

### 4.6.2 Network similarity

SimiNet [39] is a novel method for quantifying similarity between brain networks, which not only considers the edge weights of networks, but also notes the differences in spatial locations of network nodes and weighs the corresponding costs of their differences, including node replacement, node insertion, and node deletion. The SimiNet between the two matrices is defined as follows,

$$d = 1 \big/ (1 + \sum_{k=1}^{N_e - 1} \sum_{t=k+1}^{N_e} \left| c_{t,k}^1 - c_{t,k}^2 \right|) \tag{7}$$

where $c_{k,t}$ denotes the edge weight between the node pair $k$ and $t$ in the functional connectivity matrix, $N_e$ is the number of nodes. Here, the SimiNet $\in [0,1]$, 0 and 1 indicate the two connectivity matrices are dissimilar and completely identical, respectively.

## 4.7 Statistical analysis

To compare the statistical differences of either PaLOSi values or SE values between two groups of interest, we first applied the Lilliefors test and Levene test to assess the normality and homoscedasticity, respectively. When the values failed to meet the assumptions of parametric tests, the non-parametric Mann-Whitney U test was employed to evaluate the statistical differences between groups and report the p values. Using Pearson correlation to measure the degree of correlation between two features and reporting its significance.